\documentclass[conference,letterpaper]{IEEEtran}

\usepackage{amsmath,amssymb}
\usepackage{array}
\usepackage{balance}
\usepackage{booktabs}
\usepackage{cite}
\usepackage{graphicx}
\usepackage{multirow}
\usepackage{url}
\usepackage{xcolor}
\usepackage{cite}
\usepackage{amsmath,amssymb,amsfonts,mathtools}
\usepackage{algorithm}
\usepackage{algpseudocode}
\usepackage{array}
\usepackage{graphicx}
\usepackage{textcomp}
\usepackage{url}
\usepackage{gensymb}
\usepackage{nicefrac}
\usepackage[caption=false,font=footnotesize]{subfig}

\IEEEoverridecommandlockouts

\graphicspath{{Fig/}}
\newcommand{\paperfigure}[2]{%
  \IfFileExists{Fig/#1}{%
    \includegraphics[width=\linewidth]{Fig/#1}%
  }{%
    \fbox{\parbox[c][#2][c]{0.94\linewidth}{\centering
      \textbf{Figure space reserved}\\[2pt]
      \texttt{\detokenize{Fig/#1}}}}%
  }%
}

\newcommand{\uit}{u_{\mathrm{IT}}}
\newcommand{\ppcc}{P_{\mathrm{PCC}}}
\newcommand{\qpcc}{Q_{\mathrm{PCC}}}
\newcommand{\pups}{P_{\mathrm{UPS}}}
\newcommand{\qups}{Q_{\mathrm{UPS}}}
\newcommand{\vpos}{V_{1}}
\newcommand{\vneg}{V_{2}}

\title{Physics-Guided Multi-Output Surrogate Modeling and Electrical Realization of a Centralized-UPS Data Center}

\title{Physics-Guided Multi-Output Surrogate Modeling of a Centralized-UPS Data Center}

\author{\IEEEauthorblockN{
 Jesus D. Vasquez-Plaza*, Sunil Subedi*, Jin Dong, Joao Pereira Pinto}
\vspace{-1em}

\thanks{Preprint submitted to IEEE PESGM 2027. This work has been submitted
to the IEEE for possible publication. Copyright may be transferred without
notice, after which this version may no longer be accessible.}
\thanks{
This manuscript has been authored by UT-Battelle, LLC, under contract DE-AC05-00OR22725 with the US Department of Energy (DOE). The US government retains and the publisher, by accepting the article for publication, acknowledges that the US government retains a nonexclusive, paid-up, irrevocable, worldwide license to publish or reproduce the published form of this manuscript or allow others to do so for US government purposes. DOE will provide public access to these results of federally sponsored research in accordance with the DOE Public Access Plan (http://energy.gov/downloads/doe-public-access-plan).
}
\thanks{
The Authors are from the Electrification and Energy Infrastructures Division, Oak Ridge National Laboratory, Oak Ridge, TN, USA,
(email:  {vasquezplazj, subedis, dongj, pintoj}@ornl.gov). * : Corresponding Authors}}

\begin{document}
\maketitle


\begin{abstract}
Data centers supplied through centralized uninterruptible power supplies (UPSs) are converter-dominated loads whose grid response reflects IT demand, power-conversion controls, energy storage, cooling, and site-support dynamics. This paper develops and electrically realizes a causal multi-output surrogate for a nominal 1-MW, 0.48-kV simulated data center. Four physics-informed terminal features---normalized IT demand, positive-sequence voltage deviation, voltage-unbalance ratio, and frequency deviation---are used to predict active and reactive power at both the point of common coupling (PCC) and UPS output. Six causal surrogate families are evaluated using separate training, validation, and held-out trajectories. A ridge-regularized low-order nonlinear autoregressive model with exogenous inputs (NARX) achieves the lowest validation and test errors, with mean four-output errors of 0.1959\% and 0.1441\% on a 1-MVA base, respectively. The selected surrogate is then coupled to a three-phase electrical interface in MATLAB/Simulink, where predicted PCC active power controls a variable resistive load and predicted reactive power is realized through controlled current sources synchronized to the PCC voltage. Across two 41-s AI workload profiles, including an additional Pipeline Parallelism trajectory, the electrical realization tracks the NARX commands with active- and reactive-power RMSEs of 0.604~kW and 0.353~kvar for Evaluation Profile A and 1.539~kW and 0.442~kvar for Evaluation Profile B, respectively. The results demonstrate a compact terminal surrogate that can
reproduce both predicted power trajectories and their corresponding
three-phase electrical behavior under measured PCC excitation.
\end{abstract}



\section{Introduction}
Artificial-intelligence computing is making data centers larger, more concentrated, and more dynamic. Unlike conventional aggregate loads, the response observed by the grid is shaped by cascaded converters, dc-link storage, rack power supplies, cooling drives, protection logic, and correlated workload changes \cite{song_gridchip,sun_emulator,zhao_resonance}. Component and impedance models have shown how converter controls and internal distribution networks can create multiple time scales and resonance paths \cite{sun_part1,sun_part2}. Centralized-UPS studies further show that rectifier, dc-link, and inverter dynamics can materially affect power transfer under fast IT-load changes and weak-grid conditions \cite{vasquez_naps,vasquez_tsg}. Recent data-center load models for system-stability studies likewise emphasize the importance of UPS dynamics, cooling loads, and rapidly varying AI workloads \cite{jimenez2026data}.

Measurements of accelerator clusters and composite transmission models show that a static demand cannot represent workload, delivery, and facility dynamics \cite{li_ai,gyang_dynamic}. Detailed transient and component-informed models are indispensable for mechanism studies, but they require internal parameters that may be proprietary and can be expensive to reproduce in repeated or large-scale power-system studies. Measurement-driven surrogates provide a complementary behavioral representation when only terminal signals are available \cite{subedi_overview,song_gridchip}. Their credibility, however, depends on causal inputs, stable recursive operation, trajectory-level separation, and an explicit validity domain; random splits of adjacent samples from a single transient are therefore insufficient evidence of transfer to a new operating trajectory.

\begin{figure*}[h]
\centering
\includegraphics[width=0.85\textwidth]{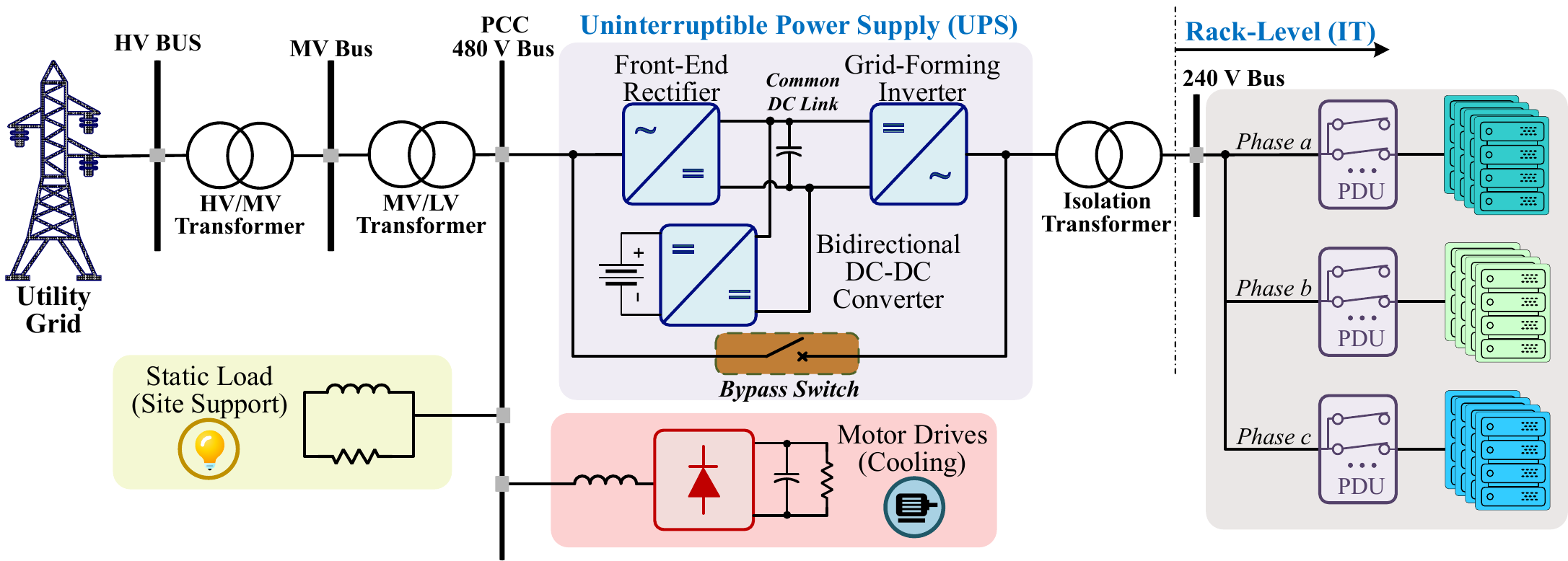}
\vspace{-1em}
\caption{Simulated data-center boundary and measurement locations. The 0.48-kV PCC meter is upstream of the UPS/IT and facility-load branches and therefore measures the complete facility. The UPS power meter is between the UPS output and the directly connected IT load; no downstream 480/240-V transformer is present. Positive active and reactive power point from the grid or UPS toward the load (improvised and adapted from~\cite{vasquez_naps}).}
\label{fig:topology}
\vspace{-1.5em}
\end{figure*}

The audited literature includes detailed converter models, positive-sequence reductions, impedance methods, and general data-driven converter-identification techniques. However, it does not provide a common-protocol comparison of causal black-box and physics-guided surrogates for the present boundary: complete-facility PCC power together with UPS-output branch power, learned from direct MW/MVAr channels and evaluated on distinct workload trajectories. In addition, surrogate-model studies commonly stop at offline signal prediction rather than demonstrating an explicit three-phase electrical realization of the predicted terminal powers.

This paper therefore makes three contributions:
\begin{itemize}
\item It defines a leakage-resistant, multi-output terminal interface for a centralized-UPS data center: normalized IT demand and PCC-voltage features are model inputs, PCC and UPS active/reactive powers are outputs, and PCC current and dc-link voltage are withheld diagnostics.
\item It compares six causal surrogate families using Profile~2 for fitting, Profile~1 for model-family selection, and Profile~3 only for held-out assessment, resulting in a compact low-order NARX with superior validation and test accuracy.
\item It develops a three-phase electrical realization of the selected NARX in MATLAB/Simulink, where predicted PCC active power is reproduced through a variable resistive branch and predicted reactive power through synchronized controlled current sources, and validates the resulting electrical tracking quantitatively.
\end{itemize}

\section{System Boundary and Identification Data}

\subsection{Facility and Measurements}

Fig.~\ref{fig:topology}  shows the simulated centralized-UPS data center and the measurement boundary used for surrogate identification. The nominal facility comprises 769 kW of IT capacity, 180.8 kW of cooling demand, and 49.9 kW of site-support load, corresponding to 999.7 kW of total rated active power and a nominal power-usage effectiveness (PUE) of 1.30. The IT aggregate is connected directly to the UPS output.  Consequently, $P_{\mathrm{PCC}}$ and $Q_{\mathrm{PCC}}$ represent the complete facility response, while $P_{\mathrm{UPS}}$ and $Q_{\mathrm{UPS}}$ quantify the power delivered from the UPS to the IT branch.

Each simulated trajectory records time; three-phase PCC-voltage magnitudes and angles; three-phase PCC-current magnitudes and angles; PCC active and reactive power; IT active power; UPS-output active and reactive power; and dc-link voltage. Voltage and current magnitudes are stored in per unit, phase angles in degrees, powers in MW/MVAr, IT demand in MW, and dc-link voltage in volts.


Let $\vpos$ and $\vneg$ denote the positive- and negative-sequence PCC-voltage phasors. The causal surrogate interface is
\begin{align}
u_k & = \left[
u_{\mathrm{IT},k},\ |V_{1,k}|-1,\
\frac{|V_{2,k}|}{\max(|V_{1,k}|,\epsilon)},\ \Delta f_k
\right]^{\!T}, \nonumber\\
y_k & = [P_{\mathrm{PCC},k},\ Q_{\mathrm{PCC},k},\
P_{\mathrm{UPS},k},\ Q_{\mathrm{UPS},k}]^{T},
\label{eq:io}
\end{align}
where $\uit=P_{\mathrm{IT}}/0.769$~MW and $\Delta f$ is obtained from a causal difference of the unwrapped positive-sequence angle. Positive power denotes consumption from the upstream source.


\subsection{Preprocessing and Trajectory Separation}

Table~\ref{tab:protocol} summarizes the frozen identification protocol. Three 41-s dynamic trajectories are used for model development. The original 50-kHz signals are processed by a causal sixth-order 180-Hz anti-aliasing filter and decimated to 500~Hz, yielding a 2-ms model time step. The four target powers are subsequently processed by a causal fourth-order 20-Hz low-pass filter. The first second of each trajectory is discarded to remove common startup behavior.
\vspace{-2em}

\begin{table}[h]
\caption{Frozen Identification Protocol}\vspace{-1em}
\label{tab:protocol}
\centering
\scriptsize
\renewcommand{\arraystretch}{1.08}
\setlength{\tabcolsep}{3pt}
\begin{tabular}{p{0.26\columnwidth}p{0.62\columnwidth}}
\toprule
Item & Setting \\
\midrule
Source records & Three 41-s trajectories at 50~kHz; first 1~s removed \\
Anti-aliasing & Causal sixth-order 180-Hz low-pass filter \\
Model rate & 500~Hz ($T_s=2$~ms) \\
Targets & Four powers causally low-pass filtered at 20~Hz \\
Training & Profile~2; first 75\% fit, last 25\% internal tuning \\
Selection & Profile~1 external validation \\
Final assessment & Profile~3 held-out test \\
Inputs & $\uit$, $|\vpos|-1$, $|\vneg|/|\vpos|$, and $\Delta f$ \\
Outputs & $\ppcc$, $\qpcc$, $\pups$, and $\qups$ \\
Withheld diagnostics & PCC current and dc-link voltage \\
Scaling & Estimated from Profile~2 only \\
\bottomrule
\end{tabular}
\vspace{-1em}
\end{table}

The trajectories are separated by operating profile rather than randomly by sample. Profile~2 is used for model fitting, with its first 75\% used for parameter estimation and the remaining 25\% for chronological internal tuning. Profile~1 is reserved for external model-family selection, while Profile~3 is used only for the final held-out assessment. Input and output scaling parameters are estimated exclusively from Profile~2.

\begin{figure}[h]
\centering
\includegraphics[trim={0.1cm 0.1cm 0.1cm 0.1cm},clip, width=0.9\columnwidth]{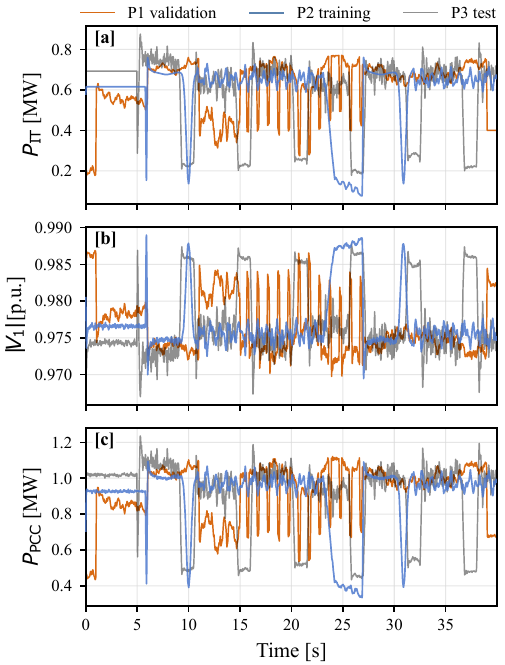}
\vspace{-1em}
\caption{Operating trajectories used for model development: Profile~2 for training, Profile~1 for external validation and model selection, and Profile~3 for held-out testing. 
}
\label{fig:profiles}
\vspace{-1.5em}
\end{figure}

As illustrated in Fig.~\ref{fig:profiles}, the three profiles contain distinct IT-demand and voltage trajectories. Profile~3 reaches approximately 1.16~p.u. IT demand, exceeding the approximately 1.0-p.u. maximum observed during training and thereby providing limited workload extrapolation. In contrast, the positive-sequence PCC-voltage magnitude remains within a comparatively narrow range of approximately 0.966--0.991~p.u.; therefore, the dataset does not establish generalization to faults or wide-range voltage disturbances.

\section{Surrogate Models and Model Selection}

\subsection{Candidate Models and NARX Formulation}

Six causal model families are evaluated using the common terminal interface and preprocessing defined in Section~II: a physics-based first-order model, a grey-box $P$--$Q$ state-space model, a physics-informed recurrent reservoir, a low-order NARX, a neural NARX, and a causal convolutional/temporal convolutional network (TCN). The physics-based and grey-box models provide compact dynamic baselines, whereas the reservoir and neural models introduce progressively greater nonlinear representation capacity. All model families are trained and selected using the same trajectory-level protocol to ensure that performance differences arise from model structure rather than data partitioning.

The selected low-order NARX augments the exogenous input vector with $u_{\mathrm{IT}}^2$ and the interaction $u_{\mathrm{IT}}(|V_1|-1)$. In standardized coordinates, its free-running recurrence is
\begin{equation}
\widetilde y_k
=
b+
\sum_{\ell\in\mathcal L_u}
B_{\ell}\widetilde{\boldsymbol\phi}_{k-\ell}
+
\sum_{\ell\in\mathcal L_y}
A_{\ell}\widetilde{ y}_{k-\ell},
\label{eq:narx}
\end{equation}
where $\mathcal L_u=\{0,1,2,5,10\}$, and $\mathcal L_y=\{1,2,5,10\}$.

The model uses a ridge coefficient of $10^{-4}$, resulting in 46 regressors per output and a maximum dynamic lag of 20ms. During evaluation, predicted rather than measured outputs are recursively fed back. A measured steady-state history initializes the first 20 samples (40ms), and the first 0.5s of each evaluated trajectory is excluded from the reported scores.

\subsection{Evaluation Metric}

For trajectory $s$, the primary multi-output score is defined on a common 1-MVA facility base as
\begin{equation}
E_s=
\frac{100}{4S_b}
\sum_{j=1}^{4}
\mathrm{RMSE}_{j,s},
\qquad S_b=1~\mathrm{MVA},
\label{eq:common_metric}
\end{equation}
with\vspace{-0.5em}
\begin{equation}
\mathrm{RMSE}_{j,s}
=
\sqrt{
\frac{1}{N_s}
\sum_{k=1}^{N_s}
\left(y_{j,k}-\widehat y_{j,k}\right)^2
}.
\end{equation}

Using a common physical base gives equal weighting to the four predicted power channels and avoids overemphasizing the small-span $Q_{\mathrm{UPS}}$ signal.


\section{Electrical Realization of the NARX Surrogate}

The selected NARX predicts four terminal quantities,
$P_{\mathrm{PCC}}$, $Q_{\mathrm{PCC}}$, $P_{\mathrm{UPS}}$, and
$Q_{\mathrm{UPS}}$. For grid-facing electrical realization, the predicted PCC powers are used to construct a balanced three-phase load connected at the data-center PCC. The active- and reactive-power components are realized separately to avoid introducing unnecessary dynamic states into the equivalent load.

\begin{figure}[t]
    \centering
    \includegraphics[width=0.86\columnwidth]{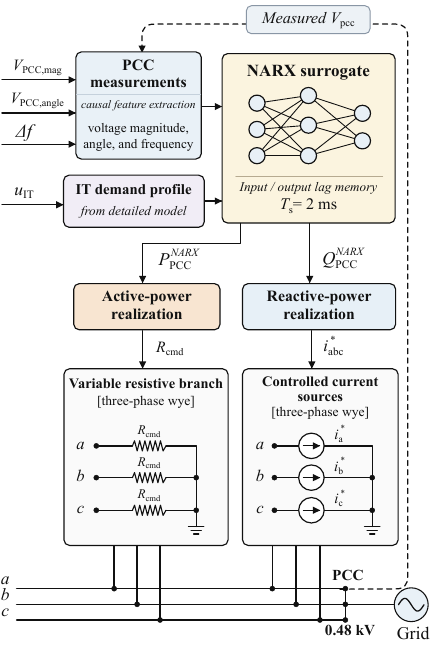}
    \vspace{-1em}    \caption{Electrical realization of the selected NARX surrogate at 0.48kV PCC.}
    
    \label{fig:electrical_realization}
    \vspace{-1em}
\end{figure}

\subsection{Active-Power Realization}

The predicted active-power command $P_{\mathrm{PCC}}^{\mathrm{NARX}}$ is reproduced using a balanced variable resistive branch. For a three-phase wye-connected load, the required per-phase resistance is
\begin{equation}
R_{\mathrm{tar}}
=
\frac{V_{LL}^{2}}
{P_{\mathrm{PCC}}^{\mathrm{NARX}}},
\label{eq:Rtarget}
\end{equation}
where $V_{LL}$ is the measured PCC line-to-line RMS voltage. To avoid an abrupt connection at startup, the resistance is ramped in the conductance domain rather than directly in resistance:
\begin{equation}
G_{\mathrm{cmd}}
=
(1-\alpha)G_{\mathrm{off}}
+
\alpha G_{\mathrm{tar}},
\qquad
R_{\mathrm{cmd}}=G_{\mathrm{cmd}}^{-1},
\label{eq:Gsoftstart}
\end{equation}
where $G_{\mathrm{tar}}=1/R_{\mathrm{tar}}$,
$G_{\mathrm{off}}=1/R_{\mathrm{off}}$, and
$\alpha\in[0,1]$ is the soft-start factor. Voltage and resistance limits are applied in implementation to prevent unrealistically large commands during low-voltage or initialization conditions.

\subsection{Reactive-Power Realization}

The reactive-power command $Q_{\mathrm{PCC}}^{\mathrm{NARX}}$ is realized using three controlled current sources. The required RMS phase-current magnitude for a balanced reactive load is
\begin{equation}
I_Q
=
\frac{Q_{\mathrm{PCC}}^{\mathrm{NARX}}}
{\sqrt{3}V_{LL}}.
\label{eq:IQ}
\end{equation}

The PCC voltage is synchronized through a phase-locked loop (PLL). With the implemented convention $v_a\propto\cos(\theta_{\mathrm{PLL}})$, positive inductive reactive-power absorption is obtained by commanding currents that lag the phase voltages by $90^\circ$:
\begin{equation}
\begin{aligned}
i_a^\ast &= \sqrt{2}I_Q\sin(\theta_{\mathrm{PLL}}),\\
i_b^\ast &= \sqrt{2}I_Q\sin(\theta_{\mathrm{PLL}}-2\pi/3),\\
i_c^\ast &= \sqrt{2}I_Q\sin(\theta_{\mathrm{PLL}}+2\pi/3).
\end{aligned}
\label{eq:Iabc}
\end{equation}

The same soft-start factor is applied to the reactive-power command before current-reference generation. In the implemented case, the electrical interface is enabled at $t=2$~s with $\dot{\alpha}=5~\mathrm{s}^{-1}$, reaching full command at $t=2.2$~s. The NARX state is reset for one 2-ms model step beginning at $t=1.8$~s.

Fig.~\ref{fig:electrical_realization} summarizes the resulting interface. The present realization uses PCC voltage measurements from the detailed simulation to provide the surrogate voltage features and synchronization angle; therefore, it evaluates electrical realization of the NARX-predicted terminal powers under measured PCC excitation rather than autonomous closed-loop replacement of the detailed data-center model.

\section{Results and Discussion}

\subsection{Surrogate Selection and Held-Out Performance}

Table~\ref{tab:results} summarizes the common-base model-selection results. The low-order NARX is selected solely from Profile~1, where its 0.1959\% validation error is 27.4\% below the physics-informed reservoir runner-up. The selection transfers to the held-out Profile~3: the NARX achieves a mean four-output error of 0.1441\%, followed by the TCN at 0.2849\% and the reservoir at 0.2866\%. The NARX fit-and-selection time is 0.77~s, compared with 38.64~s for the neural NARX and 575.53~s for the TCN in the same workstation run. These values are machine-specific rather than portable benchmarks, but the approximately $50\times$ and $750\times$ differences expose the accuracy--complexity tradeoff. The reservoir is marginally faster at 0.69~s, but has 38\% larger validation error and almost twice the held-out error.
\vspace{-1.5em}
\begin{table}[h]
\caption{Mean Four-Output Error and Fit Cost}\vspace{-1em}
\label{tab:results}
\centering
\scriptsize
\setlength{\tabcolsep}{3.2pt}
\begin{tabular}{lrrrr}
\toprule
Model & Train & Val. & Test & Time \\
 & \multicolumn{3}{c}{RMSE [\% of 1-MVA base]} & [s] \\
\midrule
Physics first-order & 0.209 & 0.484 & 0.411 & 3.52 \\
\textbf{Low-order NARX} & \textbf{0.097} & \textbf{0.196} & \textbf{0.144} & 0.77 \\
Grey-box $P$--$Q$ SS & 0.309 & 0.637 & 0.585 & 19.31 \\
PI reservoir SS & 0.137 & 0.270 & 0.287 & \textbf{0.69} \\
Neural NARX & 0.309 & 0.365 & 0.304 & 38.64 \\
Causal CNN/TCN & 0.123 & 0.357 & 0.285 & 575.53 \\
\bottomrule
\end{tabular}
\vspace{-1em}
\end{table}

\begin{figure}[h]
\centering
\vspace{-1em}
\includegraphics[width=0.85\columnwidth]{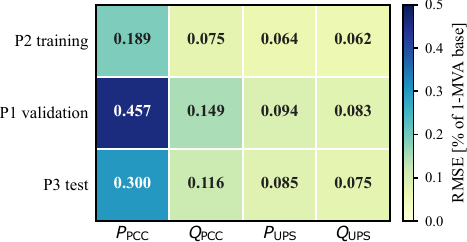}
\vspace{-1em}
\caption{Surrogate-model performance. (a)--(d) Measured and low-order-NARX-predicted powers for held-out Profile~3. (e) Channel-wise RMSE for the training, validation, and test trajectories, expressed as percentages of the 1-MVA facility base. Insets in (a)--(d) magnify the common 20--25-s interval.}
\label{fig:test_trajectorie}
\end{figure}

Fig.~\ref{fig:test_trajectorie} combines the held-out Profile~3 trajectories with the channel-wise RMSE summary across training, validation, and test profiles. For Profile~3, the test RMSE values are 0.00300~MW for $P_{\mathrm{PCC}}$, 0.00116~MVAr for $Q_{\mathrm{PCC}}$, 0.000855~MW for $P_{\mathrm{UPS}}$, and 0.000750~MVAr for $Q_{\mathrm{UPS}}$. These correspond to 0.300\%, 0.116\%, 0.085\%, and 0.075\% of the 1-MVA facility base, respectively. Active-energy biases are $-0.087\%$ at the PCC and $+0.0037\%$ at the UPS output.
\vspace{-1em}

\begin{figure}[h]
\centering
\includegraphics[width=0.9\columnwidth]{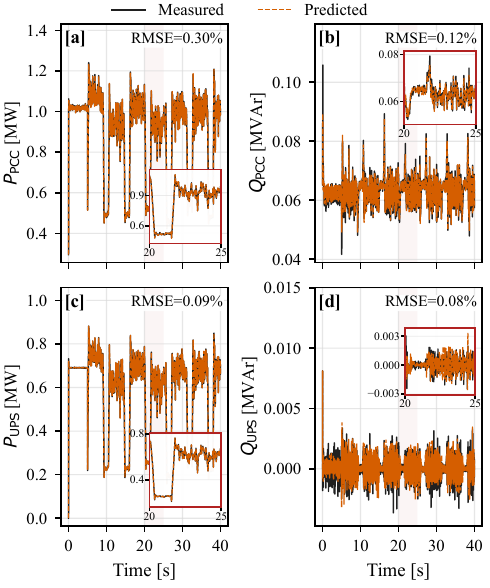}
\label{fig:test_trajectories}
\vspace{-1em}
\end{figure}

The most visible mismatch occurs during the fastest $P_{\mathrm{PCC}}$ transitions, while the UPS-output active-power trajectory is reproduced more closely. The apparently low $R^2$ of $Q_{\mathrm{UPS}}$ is not used as a headline statistic because that signal is centered near zero and exhibits very small variance; its absolute test RMSE is only 0.75~kVAr. Conversely, $P_{\mathrm{PCC}}$ dominates the selected model's error and therefore remains the most relevant channel for further model refinement.



\subsection{Electrical-Realization Performance}
\begin{figure}[t]
    \centering
    \includegraphics[trim={0cm 0cm 0cm 0cm},clip,width=\columnwidth]{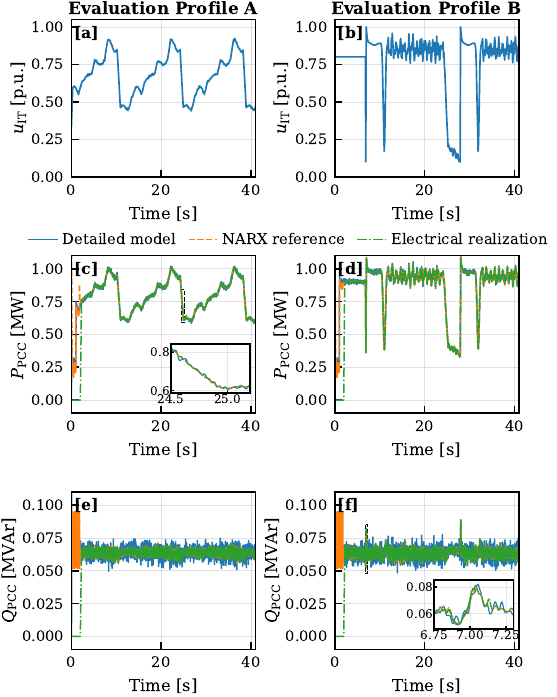}
    \vspace{-2em}
    \caption{Electrical-realization evaluation under two 41-s AI
workload trajectories. (a) Evaluation Profile A and (b) Evaluation Profile B. (c)--(d) PCC active power and (e)--(f) PCC reactive power for Profiles A and B, respectively. Power panels compare the detailed-model, NARX-reference, and electrical-realization responses; insets magnify representative intervals.}
    \label{fig:electrical_tracking}
    \vspace{-1em}
\end{figure}

To assess the electrical interface under distinct AI workload
dynamics, two 41-s trajectories are considered. Evaluation
Profile A corresponds to an additional Pipeline Parallelism
workload not used for surrogate identification, whereas Evaluation
Profile B corresponds to the DGX-H100 rack workload used during
surrogate training from the EdgeTunePower AI Load Profile Generator\cite{EdgeTunePower2026}. Fig.~\ref{fig:electrical_tracking} shows the two workload trajectories
together with the detailed-model, NARX-reference, and electrically
realized $P_{\mathrm{PCC}}$ and $Q_{\mathrm{PCC}}$ responses.

Because the electrical interface reaches full command at
$t=2.2$~s, the realization-tracking metrics are computed over
$t\geq2.2$~s. For Evaluation Profile A, the electrical realization
tracks the NARX commands with RMSEs of 0.604~kW for active
power and 0.353~kvar for reactive power. For Evaluation Profile B,
the corresponding RMSEs are 1.539~kW and 0.442~kvar.
Expressed on the common 1-MVA facility base, these errors are
0.0604\% and 0.0353\% for Profile A and 0.1539\% and
0.0442\% for Profile B, respectively. Thus, even for the more
rapidly varying Profile B workload, the additional error introduced
by the electrical interface remains below 0.154\% of the facility
base for active power and 0.045\% for reactive power.

The variable-resistance branch therefore reproduces the
NARX-predicted active-power trajectory, while the synchronized
controlled-current sources reproduce the predicted reactive-power
demand across both evaluation trajectories. The present evaluation
uses PCC excitation obtained from the detailed data-center
simulation and therefore validates electrical realization of the
predicted terminal powers under measured PCC excitation rather
than autonomous closed-loop replacement of the detailed model.

\section{Conclusion}
This paper presented a physics-guided multi-output surrogate and electrical realization for a centralized-UPS data center. Among six candidate models, the low-order NARX provided the best accuracy--complexity tradeoff, achieving validation and held-out test errors of 0.1959\% and 0.1441\%, respectively, with only 0.77~s of fitting and model-selection time. The selected surrogate was mapped to a three-phase electrical interface using a variable-resistance branch for active power and synchronized controlled current sources for reactive power. Across two 41-s AI workload profiles, the electrical realization maintained NARX-to-electrical tracking errors below 0.154\% of the 1-MVA facility base for active power and below 0.045\% for reactive power. These results demonstrate that the proposed approach can preserve
the NARX-predicted terminal power behavior through a three-phase
electrical representation while retaining the low computational
complexity of the surrogate. Future work will extend the approach toward autonomous closed-loop operation and broader grid-disturbance validation.

\balance

\bibliographystyle{IEEEtran}
\bibliography{references}

\end{document}